\documentclass[11pt]{article} 
\usepackage{bbold}
\usepackage{amssymb,amsmath,bbm,cite}
\usepackage[all]{xy}
\usepackage{graphics}
\usepackage{shadow}

\long\def\@makefntext#1{\parindent 0cm\noindent
\hbox to 1em{\hss$^{\@thefnmark}$}#1}

\def\be{\begin{equation}}
\def\ee{\end{equation}}
\def\beq{\begin{equation}}
\def\eeq{\end{equation}}
\def\bea{\begin{eqnarray}}
\def\eea{\end{eqnarray}} 
\def\beqa{\begin{equation}\begin{array}{l}}
\def\eeqa{\end{array}\end{equation}}

\def\eqn#1{(\ref{#1})}

\def\eqref#1{eq.~(\ref{eq:#1})}

\def\half{\mbox{\small{$\frac{1}{2}$}}}

\def\nn{\nonumber}

\newcommand{\SL}{{\rm SL}}

\newcommand{\ssp}{{\scriptscriptstyle +}}
\newcommand{\ssm}{{\scriptscriptstyle -}}
\newcommand{\sspm}{{\scriptscriptstyle \pm}}

\def\RR{\mathbb R}

\begin{document}


\vspace{.8cm}
\setcounter{footnote}{0}
\begin{center}
{\Large{\bf 
Topologically Massive AdS Gravity}
    }\\[10mm]

{\sc S.\ Carlip$^*$, S.\ Deser$^\ddag$,
A.\ Waldron$^\dag$, and D.~K.~Wise$^\dag$
\\[6mm]}

{\em\small  
$^*$Department of Physics,
University of California, Davis, CA 95616,
USA\\ {\tt carlip@physics.ucdavis.edu}}\\[5mm]
{\em\small  
$^\ddag$California Institute of Technology, Pasadena, CA 91125 and\\
Brandeis University, Waltham,
MA 02454, 
USA\\ {\tt deser@brandeis.edu}}\\[5mm]
{\em\small  
$^\dag$Department of Mathematics,
University of California, Davis, CA 95616,
USA\\ {\tt wally,derek@math.ucdavis.edu}}\\[5mm]


\bigskip

\bigskip

{\sc Abstract}\\
\end{center}

{\small
\begin{quote}

We analyze $(2+1)$-dimensional gravity with a Chern--Simons term and a 
negative cosmological constant,  primarily at the weak field level.  The full 
theory is expressible as the sum of  two higher derivative $\SL(2,\RR)$ 
``vector'' Chern-Simons terms, while the physical bulk degrees of freedom 
correspond to a single massive scalar field, just as for $\Lambda=0$.
The interplay of $\Lambda$ and the mass parameter $\mu$ can  be 
studied, and any  physical mass---including the conformal value with 
null propagation---is  accessible by tuning $\mu$.  The single bulk mode 
yields a complete set of normalizable positive energy wave packets, as 
long as one chooses the  usual, ``wrong" sign of $G$ necessary to connect 
smoothly with the known $\Lambda=0$ limit.   The chiral Chern--Simons 
coupling leads to gauge invariant linearized curvatures propagating with 
chirality-dependent masses.  Linearized metric fluctuations have a finite 
asymptotic Fefferman--Graham expansion about the Poincar\'e metric for any 
mass value greater or equal to a ``critical'' one, where various amusing 
effects appear, such as vanishing of one of the two ``vector" Chern-Simons 
terms and an equivalence between tensor and vector excitations.  We also find 
a set of chiral, pp-wave metrics that exactly solve the full nonlinear equations. 

\end{quote}}

\section{Introduction}

\noindent
We report some novel properties of a currently popular 
extension~\cite{Deser:2003vh,Kraus,Solodukhin,Strom,Grumiller:2008qz} of $(2+1)$-dimensional 
topologically massive gauge theories~\cite{Deser:1982vy}.  The latter are
described by vector (Abelian or Yang-Mills) or tensor (Einstein) actions
supplemented by Chern-Simons terms, while the generalizations adjoin an
anti-de Sitter (AdS) background for vectors and a cosmological constant for gravity.  Just as 
for four-dimensional massive tensor models~\cite{Deser:1983tm,Deser:2001pe}, 
the $(m^2, \Lambda)$ plane displays unexpected richness, beyond the surprises already found in
the $\Lambda=0$ models~\cite{Deser:1982vy}.  The technology
is sufficiently involved that we only provide and motivate some highlights of the 
emergent properties, and do so primarily for weak field excitations of  gravity 
about the AdS vacuum. Details may be found in~\cite{us}. 

Two effects are particularly striking.  The first is that although there is one bulk 
excitation for all mass values, suitable tuning can yield null propagation 
at finite mass, even without new gauge invariances of the sort that appear
for massive higher spins in higher dimensions~\cite{Deser:1983tm,Deser:2001pe}. 
 Even more surprisingly, the 
various curvature components of this excitation propagate with different, 
chirality-dependent masses. 

We shall first survey scalars and vectors in AdS$_3$, showing that a vector 
field  has a single, gauge-invariant, massive bulk mode.  We display the  
chirality-dependent masses of the individual field strength components,
and show how $(m^2, \Lambda)$ tuning can effect null cone propagation. 
We next analyze the more complex gravity model, which displays the same 
pattern: a single physical gauge-invariant ``scalar" mode whose corresponding  
curvature components carry chirality-dependent masses.  The bulk excitations 
consist of a complete set of  localized wave packets, all with positive energy 
(with the same overall sign choice as for $\Lambda=0$) and good asymptotic
behavior.  Separately, there are exact pp-wave chiral solutions.  Both sets of 
solutions are present for all masses, modulo some fine print below certain mass 
bounds.  Finally, we cast our model into ``pure Chern--Simons" form and note 
some formal properties at critical mass values.

\section{AdS$_3$ and Scalars}
 
\noindent 
We denote the AdS metric $g_{\mu\nu}$ and the full dynamical metric 
$\gamma_{\mu\nu}$, with signature $({-}{+}{+})$.  In our conventions, 
the Ricci tensor is $R_{\mu\nu}= \partial_\rho\Gamma^\rho_{\mu\nu} 
+ \dots = 2\Lambda g_{\mu\nu}$; the Riemann tensor is its double dual.  We 
choose units $\Lambda=-1$.    The Poincar\'e patch, 
\be
ds^2 = g_{\mu\nu} dx^\mu dx^\nu = (2dx^+ dx^- + dz^2)/z^2\, ,\label{Poincare}
\ee
covers only part of AdS, but the remainder is easily reached by inversions.  We often
make the lightlike time choice $\tau=x^+$.  

Consider first a standard (positive-energy)  massive scalar, whose unimproved 
action reduces to 
\be
\label{scalaraction}
I=\int d\tau \left\{\langle \phi,\dot\phi\rangle - \left(\frac12 [\partial_z \phi]^2
+\frac1{2z^2}\left[m^2 + \frac 34\right]\phi^2\right)dx^-dz  \right\}\, .
\ee
Here we have performed the field redefinition $\phi \equiv \frac{1}{\sqrt{z}}\varphi$, 
where $\varphi$ is the usual scalar field; the symplectic current is defined by the 
antisymmetric bracket $\langle A,B\rangle=\int dx^- dz A \partial_- B$.
With the above conventions, the field propagates according to 
\be
\Big[2\partial_-\partial_+ \, +\, \partial_{z}^2 - \frac{m^2+3/4}{z^2}\Big]\phi = 0\, . 
\label{scalareom}
\ee
At $m^2=-3/4$, this is the Minkowski wave equation; indeed, this mass corresponds 
to adding the usual conformal improvement term $\frac1{16} R \varphi^2$.  Masses 
as negative as $m^2=-1$ are permitted, that being the well-known 
Breitenlohner-Freedman bound~\cite{Breitenlohner:1982jf}.  A Fourier 
transform on $\partial_\pm$ to $\omega^2=E^2- p_x^2$ yields the Bessel equation
\be
\label{bessel}
\Big[\frac{d^2}{dz^2}+\frac1z\frac{d}{dz}
  +\omega^2-\frac{\nu^2}{z^2}\Big]\Big(\frac{\varphi}{z}\Big)=0\, ,
\quad
 \nu^2 = m^2 +1\, .
 \ee
Its solutions oscillate, and at half-integer values of $\nu$ they become the 
product of a slowly varying function times a Minkowski plane wave $\exp(ik\cdot x)$ with 
$k^2=0$, giving quasi-lightlike propagation. Indeed, the lowest value, $\nu^2=(1/2)^2$,
is  the ``improved" massless scalar. In the singular case $\omega=0$, i.e.,
$\partial_\pm \phi=0$,  chiral solutions appear; when viewed as metric  fluctuations, 
they are pp-waves, as we discuss later.  Some facts regarding the bulk 
solutions in our patch, whose boundary includes both $z=0$ and a line at $z=\infty$: 
finiteness at $z=0$ is achieved by choosing the Bessel function solution $J_\nu \sim z^\nu$ 
at the origin; to achieve finiteness at the other end requires choosing bounded wave 
packets, which we describe below.

\section{Vectors}

\noindent
The Abelian vector system has an action
\be
I=
-\frac14\int d^3x \Big\{ \sqrt{-g}\, F_{\mu\nu} g^{\mu\rho}g^{\nu\sigma}F_{\rho\sigma}
+\mu\ \varepsilon^{\mu\nu\rho} F_{\mu\nu}A_\rho\Big\}\, ,
\ee
where $\mu$  has the dimension of mass. Note that the Chern-Simons term is 
metric-independent. The model has a single, gauge-invariant, ``scalar" component
\be
\phi = \sqrt{z}\, [A_z -(\partial_z/\partial_-) A_-] = \sqrt{z}\, \partial^{-1}_-F_{-z}\, ,
\ee
assuming invertibility of $\partial_-$.  
Observe that since $\phi$ is gauge-invariant, the analysis simplifies by using light-front 
gauge $A_-=0$; this lesson is especially useful when we move on to the tensor case.
This $\phi$ obeys the scalar equation~\eqn{scalareom},
but with a shifted mass, $\nu^2=m^2+1=(\mu+1)^2$.   All values of $\mu$ obey the $m^2\geq-1$ 
bound.  Notably, the value required for 
lightlike propagation occurs at half-integer $\nu$ rather than at $\mu=0$. 

Now consider the three components $(F _{+ z}, F_{+ -}, F_{-z})$ of the field strength. 
Here we find a surprise: although the components come from a single ``scalar'' mode,
they propagate according to scalar wave equations with different masses,
$$m^2 = \left((\mu-1)^2-1\, ,\;  \mu^2-1\, ,\; (\mu+1)^2-1\right)\, .$$ 
This result is manifestly symmetric under simultaneous $\mu$ and helicity flips.   Finally, 
it is clear that the energy remains positive in AdS: since the vector Chern-Simons term
is purely topological, it does not contribute to the stress-energy tensor, which differs 
from its flat value only by an overall $z^2$ factor coming from the metric (\ref{Poincare}).

\section{Gravitons}

The complete action for our model, in terms of the dynamical metric $\gamma_{\mu\nu}$,
is
\be
I=\!\int\! d^3 x \Big\{
\!-\!\sqrt{-\gamma}\,  (R - \, 2 \Lambda)
+
\frac{1}{2\mu}\,\varepsilon^{\mu\nu\rho} \Big(\Gamma_{\mu\beta}^{\alpha}
\partial_\nu\Gamma_{\rho\alpha}^{\beta}
+\frac{2}{3}\, 
\Gamma_{\mu\gamma}^{\alpha}
\Gamma_{\nu\beta}^{\gamma}
\Gamma_{\rho\alpha}^{\beta}\Big)\Big\} \, .\label{TMG}
\ee
As in the $\Lambda=0$ model,  the sign of the Einstein part is ``wrong,'' opposite 
to that in four dimensions.  Writing $\gamma_{\mu\nu}=g_{\mu\nu} + h_{\mu\nu}$,
we find that physical quantities can be expressed at first order in terms of the  
linearized-diffeomorphism-invariant ``scalar" 
\be
\varphi\equiv h_{zz}\,  -\,  2([\partial_z+\frac1z]/\partial_-) h_{-z} 
  + ([\partial_z+\frac1z][\partial_z+\frac2z]/\partial_-^2) h_{--}\, .\label{invar}
\ee
This reduces to $h_{zz}$ in light front gauge $h_{\mu-}=0$, where the 
metric becomes
\be
ds^2 = \frac{2dx^+dx^- + dz^2 }{z^2}+(dx^+)^2 h_{++} 
  + 2 dx^+ dz h_{z+} + dz^2 h_{zz} + {\cal O}(h^2)\, .
\ee 
Defining $\phi= z^{3/2} h_{zz}$, we again find the scalar action (\ref{scalaraction}), 
but now with mass
\be
\label{tensmass}
m^2 = (\mu+2)^2 -1\, .
\ee
The analog of the ``field strength" is the linearized cosmological Einstein tensor,
\be
{\cal H}_{\rho\sigma} 
  = \left[G_{\rho\sigma}-g_{\rho\sigma}\right]_{\scriptscriptstyle\rm LINEAR}\, ,
\label{linH}
\ee
which is invariant under linearized diffeomorphisms.  As in the vector case, the components
propagate with different masses: $({\cal H}_{++},{\cal H}_{+z},{\cal H}_{zz}
= -2{\cal H}_{+-},{\cal H}_{z-},{\cal H}_{--})$ have masses 
$$m^2 = \left(
(\mu-2)^2-1\, ,\;
(\mu-1)^2-1\, , \; 
\mu^2 -1 \, , \;
(\mu+1)^2 -1\, , \;
(\mu+2)^2 -1\right)\, .
$$
The results are again invariant under simultaneous $\mu$ and chirality flips
$\mu \leftrightarrow - \mu$, $+ \leftrightarrow -$.  

All values of $\mu$ respect the $m^2\geq-1$ bound, so the excitations represent a consistent 
positive energy system.    Moreover, the Bianchi identities guarantee that every
${\cal H}_{\mu\nu}$ solution comes from a linearized metric fluctuation.   We
must still, however, check the asymptotic behavior of the perturbed geometry.
It can be shown that the appropriate Bessel function solutions of (\ref{bessel})
lead to metrics that are asymptotically AdS at $z=0$, with finite
curvature invariants and finite contractions of ${\cal H}_{\mu\nu}$ with AdS 
unit vectors.  If we demand the stronger Fefferman-Graham~\cite{Fefferman}  
conditions of finite $h_{\mu\nu}$ at $z=0$,  on the other hand, we find a 
restriction $\mu\geq1$.

More problematic is the large $z$ behavior, where $J_\nu \sim z^{- 1/2}$, with a
consequent divergence of ${\cal H}_{\mu\nu}\sim z^{+ 1/2}$ for all $\mu$. 
This means a breakdown at the $z\rightarrow \infty$ piece of the boundary of 
our Poincar\'e patch.  Fortunately, this difficulty is circumvented by a complete, 
normalizable set of  bounded  wave packets that avoid this edge. This set exists 
for all $\mu$, including the $\mu=1$ ``critical" value, so a physical bulk degree 
of freedom is always present, as we now show. 

The $h_{zz}$ modes  are
\begin{align}
h_{\omega k}(x,z,t) &= \sqrt{\frac{\omega}{4\pi E}}\,
   \frac{1}{z}J_{\mu+2}(\omega z)e^{ikx-iEt}\, ,
    \nn \\[4mm]
h^*_{\omega k}(x,z,t) &= \sqrt{\frac{\omega}{4\pi E}}\,
   \frac{1}{z}J_{\mu+2}(\omega z)e^{-ikx+iEt} \ , 
  \label{hmodes}
\end{align}
with $E=\sqrt{\omega^2+k^2}$.
A conserved bilinear current  ${\cal J}^\mu = z^4 g^{\mu\nu}
(h_1 \partial_\nu h_2 - [1\!\leftrightarrow\!2])$
defines a time-independent Klein--Gordon inner product $(\cdot,\cdot)$
with respect to which these modes are (continuum) orthonormal:
\be
(h_{\omega k},h_{\omega' k'}) = - (h^*_{\omega k},h^*_{\omega' k'}) 
   = \delta(\omega-\omega')\delta(k-k'),  \quad (h_{\omega k},h^*_{\omega' k'}) = 0 
\, .
\ee
Arbitrary superpositions  
\be
h_{zz}(x,z,t) = \int d\omega dk\left[ a(\omega,k)h_{\omega k}(x,z,t) 
   + a^*(\omega,k)h^*_{\omega k}(x,z,t)\right] 
\label{superpose}
\ee
provide general solutions with support bounded away from $z=\infty$. 
The norm 
\be
(h,h)=\int d\omega dk [ a(\omega,k)^2-a^*(\omega,k)^2]
\ee
is constant in time and of the standard Klein-Gordon form, with the usual sign 
for the negative frequency modes.

The conserved energy of the excitations can now be evaluated, giving
\begin{align}
H &=\ \frac{1}{2} \int dx\,dx\, z^3\left[ (\partial_th)^2 + (\partial_xh)^2 
   + (\partial_zh)^2 + \frac{(\mu+2)^2-1}{z^2}h^2 \right] \, \nonumber \\
   &= \ \frac{1}{2} \int dx\,dx\, z^3 \left[ h\partial_t^2h - (\partial_th)^2\right] 
   =  \int d\omega dk\, \sqrt{k^2+\omega^2}\, |a(\omega,k)|^2 \, ,
\label{energya}
\end{align}
upon (permitted) integrations by parts.  Finite energy at an initial time then
guarantees asymptotically AdS behavior at the entire boundary for all times.
These considerations are also valid  for $\mu=1$, the point at which it was 
claimed in~\cite{Strom} that ``massive gravitons'' disappear.   The authors
of~\cite{Strom} considered only eigenfunctions of the compact $\SL(2,\RR)$  
generator; our wave packet solutions, while not such eigenstates, are
nevertheless a perfectly acceptable complete set of functions (also for $\mu=1$). They require the 
``wrong" sign choice in~\eqn{TMG} for positive energy.

\section{Chern--Simons  Formulations}

Amusingly, our model~\eqn{TMG}, in dreibein $e_\mu{}^a$ rather than metric 
variables, can be cast in ``pure," but higher derivative, Chern-Simons form
\be
I_{\scriptscriptstyle\rm TMG}[e] = -\half(1-{\textstyle \frac 1\mu})I[{}^{\ssp}\! A[e]]  
  + \half(1+{\textstyle \frac 1\mu})I[{}^{\ssm}\! A[e]]\, ,
\ee
where
\be
I[A]= \frac{1}{2}\int d^3 x \;\varepsilon^{\mu\nu\rho} \left(A_\mu{}^a{}_b
\partial_\nu A_\rho{}^b{}_ a  + \frac{2}{3}\, 
A_\mu{}^a{}_c A_\nu{}^c{}_b A_\rho{}^b{}_a \right)
\ee
and
\be
{}^{\sspm}\! A_\mu{}^a{}_b[e] 
  = \omega_\mu{}^a{}_b[e] \pm \varepsilon^{a}{}_{bc} e_\mu{}^c\, .
\ee
Here $\omega[e]$ is the usual torsion-free spin connection, and the coefficients of 
the two Chern--Simons terms are the left/right central charges.  In particular, at 
$\mu=\pm1$ the action reduces to a single term (R.~Jackiw and  D.~Grumiller
have also observed this---private communication).  
The scheme is reminiscent of the
earlier ATW~\cite{AT} formulation of cosmological Einstein gravity, 
but differs because of the higher derivatives in the gravitational Chern-Simons
term.  Our model's propagating modes are due to these higher derivatives, as has long 
been understood~\cite{DesJack}.  A canonical formulation of the full theory~\eqn{TMG}, 
extending the $\Lambda=0$ analysis of~\cite{DeserXiang}, also sets to rest any issues 
regarding the excitation content for all $\mu$.  The results of our linearized analysis 
have recently been confirmed in~\cite{Grumiller:2008pr} at $\mu=1$, and an 
independent computation~\cite{CarliP} at arbitrary $\mu$ gives the same result:  
a single bulk degree of freedom occurs for all $|\mu|\in(0,\infty)$.

In addition to the reduction from two Chern-Simons terms to one at $\mu=1$,
 a quite different special result occurs at this value:  linearized topologically massive 
gravity becomes equivalent to topologically massive electrodynamics.  Solutions of the 
latter determine all gravity solutions through the equivalence 
\be
{\cal H}_{\mu\nu}= D_{(\mu} F_{\nu)}\, ,
\ee
where $F_\nu$ is the dual Maxwell field strength and $D_\mu$ is the AdS covariant 
derivative.

\section{Chiral pp-Waves}
  We noted earlier that in addition to the (non-chiral) bulk modes, there exist chiral ones; 
we sketch them here.  In contrast to the wave packets discussed above, these chiral 
states are in fact solutions of the full nonlinear field equations.  The ansatz
\be
ds^2=\frac{2dx^+(dx^-+z^{2+\gamma}h(x^+)dx^+) + dz^2}{z^2} 
\label{chiralsoln}
\ee
solves the nonlinear field equations for the three values $\gamma = (-2,0,\mu-1)$. The 
first two are also solutions of pure Einstein gravity,  and of these only $\gamma=0$ 
is asymptotically well-behaved.  The third solution is a chiral pp-wave for all values 
of $\mu\neq\pm 1$, with finite asymptotic behavior for $\mu >1$. At $\mu=1$, the 
ansatz (\ref{chiralsoln}) must be replaced by one logarithmic in $z$, which is 
asymptotically AdS but not finite at the boundary in the Fefferman--Graham sense.

In this connection, omitting details, it is worth giving a representation-theoretic 
description of the equations and solutions in terms of the $\SL(2,\RR)\times \SL(2,\RR)$ 
AdS$_3$ isometry group, with Killing vectors
\begin{alignat}{2}
& L_+=\partial_+ \, ,\qquad &&R_+=\partial_- \, , \nonumber\\
&L_0=-x^+\partial_+-\frac12 z\partial\, , \qquad  && R_0=-x^-\partial_-
  -\frac12 z\partial\, ,\label{ISOM}\\
&L_{-}=-x^+(x^+\partial_++z\partial)+\frac12 z^2 \partial_- \, ,\qquad
  &&R_{-}=-x^-(x^-\partial_-+z\partial)+\frac12 z^2 \partial_+ \, ,\nonumber
\end{alignat}
and equal left/right Casimirs related to the scalar Laplacian by
\be
\Delta_{L/R}=\{L_+,L_-\}+2L_0^2=
\{R_+,R_-\}+2R_0^2=\frac 12\ z^2\, \Big(2\partial_+\partial_-+z\, \partial \, 
  \frac 1z\,  \partial\Big)=\frac 12\Delta\, .
\ee 
The bulk solutions we have discussed so far are not chiral, and the bulk-boundary 
propagator is the intertwiner between the reducible representation of the isometry 
group generated by the Killing vectors and irreducible quasiprimary boundary fields.
The additional chiral pp-wave solutions are obtained by requiring that $L_+$ and 
$R_+$ annihilate the highest chirality curvature component ${\cal H}_{--}$ and 
that $R_+$ annihilate all other curvature fluctuations.  In this case one finds discrete 
series representations for one of the $\SL(2,\RR)$ factors and a singlet for the other, 
and therefore a chiral subsector of the theory.

\section{Conclusions}

We have surveyed some of the novel features of cosmologically extended topologically 
massive gravity, as well as those that reduce smoothly in the $\Lambda\rightarrow0$ 
limit.  Among the latter is the existence of a single massive spin 2 excitation for all 
$\mu$, which has positive energy as long as the same ``wrong" sign Einstein term is 
used as in the $\Lambda=0$ case.  The weak field solutions consist of a complete set 
of  non-chiral wave packets, along with pp-wave chiral solutions. Among the novelties 
associated with the $(m^2, \Lambda)$ plane, we found that the curvature excitations 
propagate with masses that are both shifted and internally split according to chirality; 
and that tunings exist allowing for null propagation at special  mass values. 
We also found a two-term ``pure 
Chern-Simons" expression for our action in terms of a composite, higher-derivative 
sum of torsionless spin connection and dreibein fields; at $\mu=\pm 1$, one of these 
two terms vanishes. Separately, we formulated an explicit equivalence between 
vector and tensor excitations at $\mu=1$. 

Finally, we mention the possible stability issue posed by the BTZ black hole,
which is also an exact solution to our model.  With the ``wrong" sign choice required 
to keep the bulk mode stable, the black hole has a negative mass.   Classically, however,
it is not clear that negative energy black holes can be created from positive energy 
matter or source-free excitations, and at the quantum level, we know of no instanton 
that mediates their creation.  Indeed, there may well be a superselection argument 
that excludes the BTZ metric.  One strong hint in this direction comes from the fact 
that the model has a supergravity extension (for $\Lambda<0$, as usual) which is 
necessarily  of positive energy by the usual $H=\{Q,Q^*\} >0$ relation~\cite{Deser:1984py}.

\section*{Acknowledgments}
We thank D.\ Grumiller, R.\ Jackiw, N.\ Johansson, W.\ Li, D.\ Marolf, W.\ Song, and 
A.\ Strominger for discussions. This work was supported by the National Science 
Foundation under grants 
PHY07-57190 and DMS-0636297 and the Department of Energy under grants 
DE-FG02-91ER40674 and DE-FG02-92-ER40701.

\end{document}